# Tailored Porous Electrode Resistance for Controlling Electrolyte Depletion and Improving Charging Response in Electrochemical Systems

James W. Palko,[a,b,‡] Ali Hemmatifar,[a,‡] and Juan G. Santiago[a,*]

‡ contributed equally to this work

[a] Department of Mechanical Engineering, Stanford University, Stanford, CA 94305, USA
[b] Department of Mechanical Engineering, University of California, Merced, CA 95343, USA

**Abstract**

The rapid charging and/or discharging of electrochemical cells can lead to localized depletion of electrolyte concentration. This depletion can significantly impact the system's time dependent resistance. For systems with porous electrodes, electrolyte depletion can limit the rate of charging and increase energy dissipation. Here we propose a theory to control and avoid electrolyte depletion by tailoring the value and spatial distribution of resistance in a porous electrode. We explore the somewhat counterintuitive idea that increasing local spatial resistances of the solid electrode itself leads to improved charging rate and minimal change in energy loss. We analytically derive a simple expression for an electrode resistance profile that leads to highly uniform electrolyte depletion. We use numerical simulations to explore this theory and simulate spatiotemporal dynamics of electrolyte concentration in the case of a supercapacitor with various tailored electrode resistance profiles which avoid localized depletion. This increases charging rate up to around 2-fold with minimal effect on overall dissipated energy in the system.

## 1. Introduction

Electrochemical systems are commonly used in situations which strongly benefit from the rapid storage of energy,[1] including time sensitive charging situations for electric vehicles,[2] storage of short lived energy surges as in regenerative braking,[3] and capacitive deionization (CDI) applications where increasing throughput is essential.[4] System resistances, particularly series resistances, can have a strong impact on both charge/discharge time and dissipated energy. Constant voltage charging of a simple capacitor in series with a resistance provides a useful example. For fixed resistance, the charge stored by the capacitor asymptotically approaches a maximum value following an exponential dependence with a time constant equal to the product of series resistance and capacitance ($\tau = RC$).[5] Consequently, electrochemical system design has traditionally sought to reduce all sources of resistance,[6] including solution resistance, electrode resistance, and in some cases, contact resistance between the electrodes and current collectors. Solution resistance generally dominates compared to resistance of electrode or current collector materials.[6,7]

Supercapacitors are representative of electrochemical systems designed for fast charging and high power densities, and employ strategies to minimize series resistance. The thickness of spacer layers between the electrodes is minimized to reduce solution resistance for ion transport. Likewise, electrode thickness is minimized to reduce ion transport resistance through the depth of the porous electrode. Contact resistance between electrode and current collector is minimized by a variety of collector surface preparations and electrode deposition techniques.[6] The standard design philosophy has been

**Broader context**

Electrochemical systems, including batteries and supercapacitors, are essential in energy storage. The distribution and transport of ions in electrolytes permeating porous electrodes often controls the performance of these devices. In particular, the local depletion of charge carrying ions can dramatically increase the resistance of the system, due to rapid ion removal from solution by the electrode, or due to ionic migration limitations under electric fields in the solution. This increase in local resistance can slow charging and discharging response of the system and leads to high electric fields in the device, with consequences for failure modes such as dendrite growth. Here, we consider the interplay between electronic conduction in the electrode matrix and ionic conduction in the pore space. By tailoring the spatial distribution of resistance in the electrode matrix, we show the potential to control ionic concentration evolution in the pore space, and specifically to eliminate localized electrolyte depletion. This approach holds potential for improving time response of charging (as shown here) and discharging. We here propose to accomplish this by the highly counterintuitive tactic of increasing elec-to maximize the conductivity of both current collectors and electrode materials. In this paper, we will challenge this design principle.

---

* To whom correspondence should be addressed. E-mail: juan.santiago@stanford.edu

Local solution conductivity variations can also severely impact overall performance. As an electrochemical system is charged or discharged, ions are removed from or added to solution resulting in solution conductivities which evolve in time and space. Localized depletion of the electrolyte may significantly increase the net series resistance of the cell by introducing a choke point which limits charging rate. For example, electrolyte depletion can be important in a number of battery chemistries, such as Li-ion, operating at high currents[8–10] and in supercapacitors.[11] Likewise, the ability to deplete the working solution forms the basis of CDI.[4] Significant depletion of electrolyte has a substantial effect on the charging response of the system.[12]

Depletion depends on electrolyte concentration and the charge storage capacity of the electrode. The ultimate limit for electrolyte concentration is set by its solubility in the solvent, particularly at low operating temperatures, but other constraints often further limit electrolyte concentrations including cost and decreasing conductivity with further solute addition. Many electrochemical systems use solutions based on organic solvents in order to allow larger operating voltages with negligible electrolysis.[13] Organic carbonates are popular solvent choices.[14] Lithium hexafluorophosphate ($LiPF_6$) salt is the most commonly used electrolyte in Li-ion batteries,[15] while tetraethylammonium (TEA, $(C_2H_5)_4N^+$) tetrafluoroborate (TFB, $BF_4^-$) salt is commonly used for supercapacitors.[16] Electrolyte concentrations are commonly limited to ~1 mol/L.[16,17]

Due to their high charge storage capacity, battery electrodes are usually capable of significant or complete local depletion of the electrolyte solution. This may be due to consumption of species in the electrochemical reaction (e.g. $SO_4^-$ in lead-acid cell discharge)[18] or electromigration of passive electrolyte species that do not participate in the reaction (e.g. $PF_6^-$ in Li-ion cells).[8] Commercial intercalation electrode materials for Li-ion batteries can show volumetric capacities that are many times the electrolyte concentration, exceeding 26 mol/L and 16 mol/L for cathodes and anodes, respectively, based on the total electrode volume.[19] High volumetric capacities exacerbate electrolyte depletion. Newer materials under development show capacities that are dramatically higher still. Local electrolyte distribution has been measured directly during operation using magnetic resonance imaging in supercapacitors[20] and lithium batteries.[21]

Substantial depletion can also occur with current supercapacitor electrode materials, and the potential for depletion is compounded by recent advances in electrode materials, which have dramatically increased electrode capacitance and/or pseudo-capacitance along with energy storage capability. Fang and Binder obtained specific capacitance of 160 F/g in carbon aerogels treated to improve their hydrophobic character.[22] Kim et al. showed specific capacitance of 170 F/g for aligned carbon nanotubes in acetonitrile with 1 mol/L TEA-TFB.[23] Chmiola et al. measured volumetric capacitance of 180 F/cm$^3$ for thin (~2 μm) films of carbide derived carbons in organic electrolytes.[24] Yan et al. showed specific pseudo-capacitance of 1020 F/g in 1 mol/L $LiClO_4/CH_3CN$.[25] For a symmetric capacitor, the maximum total-volume-averaged (not local) change in concentration of ions in the pore volume can be written as

$$\Delta c = \frac{CV_{max}}{2F(2p_e + p_s L_s/L_e)}, \quad (1)$$

where $C$ is volumetric capacitance, $V_{max}$ is the maximum cell voltage (split evenly in each electrode), $F$ is Faraday's constant, $p_e$ is porosity of the electrodes, $p_s$ is the porosity of the spacer, and $L_e$ and $L_s$ are respectively thickness of electrodes and spacer. As an example, for specific capacitance of 250 F/g, apparent density of 0.4 g/cm3, porosity of 80%, $V_{max}$ of 3 V, and $L_e$ of 200 μm and $L_s$ of 100 μm, the maximum concentration change ($\Delta c$) is 0.78 mol/L (i.e. 78% depletion of a standard 1 mol/L electrolyte). Newer pseudo capacitive materials would allow removal of electrolyte many times the solubility limit.

We note that the extremely large capacity of the electrodes for batteries, which provide volumetric storage of species rather than only the surface action of capacitive systems, further exacerbates localized depletion effects. Beyond the effect on charge/discharge rate, depletion has potentially important consequences for battery safety. Depletion has been implicated in the transition to dendritic lithium growth and consequent shorting in cells with lithium metal anodes.[26]

In this paper, we present what is to our knowledge a new design principle for electrochemical systems: the decrease and the tailoring of electrode conductivity to control and avoid local depletion of electrolytes and to increase charging rate of the system. We begin with an introduction to our approach based on a transmission line circuit model describing the essential coupling of distributed electrolyte and electrode conductivities. We present a numerical model based on detailed porous electrode transport theory, and use this model to study spatiotemporal dynamics of electrolyte conductivity in the spacer and electrode pores. We then derive an analytical form of electrode conductivity for uniform salt adsorption and show that a spatially variable and decreased electrode conductivity profile avoids local depletion and increases charging rate by ∼2-fold. Lastly, we show the effect of electrode conductivity on energy dissipation and the negligible effect of our tailored electrodes on overall resistive loss.

One potentially important phenomena that we do not consider here is conduction along the surface of the electrode matrix resulting from the high ionic concentration in the double layer. This effect has been shown to allow enhanced charging kinetics in porous electrodes.[27] Such conduction can help ameliorate the effect of electrolyte depletion in the bulk solution by providing an alternative ionic conduction path. However, the surface conduction path must be continuous throughout the electrode to significantly improve charging kinetics, which may limit its effect in many electrodes such as those formed by compacted powders with poor contact between particles. Furthermore, systems storing charge via Faradaic reactions, such as battery electrodes, will not generally display enhanced surface conduction.

## 2. Effect of electrode conductivity on transport dynamics

We here introduce the effect of electrode conductivity on salt adsorption dynamics using a transmission line analogy. As argued in the introduction, the specific spatial and temporal profile of ion removal from the electrolyte is intimately tied to the resistance and capacity of the rest of the electrochemical system. The response of these systems involves a complex interplay between electromigration and diffusive transport of electrolyte species throughout the porous structures of the electrodes and spacer. Figure 1 shows a useful schematic for study of temporal response and dissipation within a supercapacitor where each element represents a small thickness of the electrode or spacer (see e.g. Bisquert[28] or de Levie[29]). In response to an applied voltage, ionic current flows in the solution permeating the electrode and spacer, where it experiences non-uniform ionic resistances ($R_{ion,i}$) which depend on the local concentration of electrolyte. Electric double layers sequester ions from solution where they are balanced by electrons (or holes) in the electrode material, thus charging the local capacitive element ($C_i$) and intimately coupling the ionic current to the electronic current. The electronic current flows through the electrode to the terminal. Typically, the resistances in the electrode are represented as uniform and negligibly small compared to the electrolyte resistance.

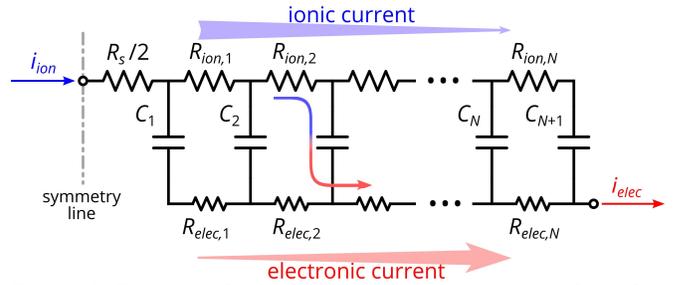

**Figure 1.** Transmission line model for a supercapacitor. Capacitor is symmetric about the indicated centerline on the left, with one electrode and half of separator represented. Current enters as ionic flux from the left and flows as ionic current through solution, represented by resistors, $R_{ion,i}$. The ionic current flows through the pore space of the electrode and charges local capacitive elements, $C_i$. The local capacitive elements can be interpreted as regions where ionic current is converted to electronic flux. Electronic current flows through the solid matrix of the electrode ($R_{elec,i}$), and total electronic current is collected at the terminal (node on bottom right). We here propose increasing values of $R_{elec,i}$ and tailoring their spatial distribution to uniformly charge the electrochemical device (avoiding ion depletion in real systems).

Indeed, as we have mentioned, electrochemical system designers often work to minimize all resistances including electrode materials, making the latter assumption accurate. To introduce our approach, we here consider appreciable and non-uniform electrode resistances $R_{elec,i}$. In traditional designs where $R_{ion,i} \gg R_{elec,i}$ for any location, the path of least resistance for the ionic current is to charge the nearest capacitive element (left-most element) favoring fast and near-spacer local conversion of ionic to electronic current that can escape to the terminal (lower right node) with minimal resistance. Under these conditions, charging of the electrode starts near the spacer and then proceeds into the depth of the electrode only when the capacitive elements near the spacer have been significantly charged. The depletion of the ions in solution will occur at the interface of the electrode and the spacer. For example, Suss et al.[30,31] and Hemmatifar et al.[32] demonstrated this effect for flow-through and flow-between capacitive deionization, respectively.

Depletion of ions accompanying charging of the electrodes can dramatically change the local solution resistance, and the resulting spatially non-uniform solution conductivity can have significant effects on the time response for charging of electrochemical systems including supercapacitors with high specific capacitance or low electrolyte concentrations. This effect has also been demonstrated at the pore scale.[33,34] As we show here, improving uniformity of charging can largely eliminate ion depletion and improve the time response of these systems.

### 3. Concept of reduced and non-uniform electrode conductivity for better performance

We here propose to improve the uniformity of charging by decreasing and tailoring the conductivity profile of the electrode. The principle is straightforward. First, we decrease the conductivity of the electrode near the spacer. This forces ionic current to penetrate deeper into the electrode before being converted to electronic current via charging of the electrode. Second, we use this principle to create a distribution of conductivities which makes charging approximately uniform and thus preventing depletion near the electrode/separator interface. As we shall show, this counterintuitive modification of decreasing electrode conductivity leads to notably improved charging time response with negligible effect on dissipated energy.

As we discuss in the next section, spatially uniform conversion of ionic to electronic current (e.g. coupling via displacement current through capacitive double layers) requires electrode conductivity similar in magnitude to the solution conductivity. Importantly, this decrease in electrode conductivity necessarily increases Ohmic losses in the electrode. However, we will show that this increased dissipation is approximately offset by the gain of avoiding Ohmic losses in the electrolyte associated with ion depletion. Hence, our approach of decreasing electrode conductivities can result in minimal overall energy penalties while substantially accelerating the charging speed.

### 4. Porous electrode model capturing a system with non-uniform electrode conductivity

We use macroscopic porous electrode (MPE) theory[35–37] to model the behavior of supercapacitors during charging and discharging. In the framework we develop here, for simplicity, we consider a binary and symmetric electrolyte with equal anion and cation diffusion constants and equal electric mobilities. This assumption results in geometric symmetry about the midplane of the cell. We treat a one-dimensional model of the cell normal to the midplane. We also consider an isothermal system at 25 °C.

Our formulation is based on MPE theory, meaning, transport equations are volume averaged. The volume averaging is performed over a scale substantially larger than pore features but small compared to the macroscopic size of the cell (in order to capture spatiotemporal variations of potential, concentration, etc.). The general form of the mass transport equation for species $i$ in a porous electrode with fixed (in space and time) porosity $p$ is

$$p \frac{\partial c_i}{\partial t} + p \nabla \cdot \boldsymbol{j}_i = s_i, \quad (2)$$

where $c_i$ is concentration of species $i$, $\boldsymbol{j}_i$ is its associated molar flux vector, and $s_i$ is the local ion source term (a signed quantity, negative during charging). The molar flux vector $\boldsymbol{j}_i$ has convection, electromigration, and diffusion contributions and can be expressed as

$$\boldsymbol{j}_i = \boldsymbol{u} c_i - \mu_i c_i \nabla \phi - D_i \nabla c_i, \quad (3)$$

where $\boldsymbol{u}$ is local flow velocity, $\phi$ is electric potential, and $D_i$ and $\mu_i$ are respectively tortuosity-corrected diffusivity and electric mobility of species $i$ in the porous electrode. We here use a simple correction for diffusivity and mobility using tortuosity as $D_i = D_{i,\infty}/\tau$ and $\mu_i = \mu_{i,\infty}/\tau$, where $\tau$ is tortuosity and $D_{i,\infty}$ and $\mu_{i,\infty}$ are respectively diffusivity and mobility of species $i$ in free solution. Further, we relate tortuosity and porosity through the Bruggeman relation[38,39] as $\tau = 1/\sqrt{p}$. We assume electroneutrality holds in the spacer and electrode pores (far from electric double layers (EDLs)). So, for a binary and symmetric electrolyte, $c_+ = c_- = c$. We model dynamics of the EDL structure with a simple Helmholtz model. This implies a constant and uniform EDL capacitance and unity charge efficiency. We further neglect bulk flow ($\boldsymbol{u} = 0$). With these assumptions, we take a localized, small-volume average of the transport equations for anions and cations and arrive at the following forms for electrodes and spacer

$$p_e \frac{\partial c}{\partial t} - p_e D_e \nabla^2 c = \frac{1}{2F} \frac{\partial \rho}{\partial t}, \quad \text{(electrode)} \quad (4)$$

$$p_s \frac{\partial c}{\partial t} - p_s D_s \nabla^2 c = 0, \quad \text{(spacer)} \quad (5)$$

where $p_e$ and $p_s$ are respectively the porosity of electrodes and spacer. Similarly, $D_e$ and $D_s$ are tortuosity-corrected diffusivity of ions in electrodes and spacer. $F$ is Faraday's constant, and $\rho$ is stored charge density (in units of Coulombs per electrode volume). We further subtract transport equations for anions and cations and arrive at the current conservation equation in the electrode

$$\frac{\partial \rho}{\partial t} = p_e \nabla \cdot \boldsymbol{i}_{ion}, \quad (6)$$

where $i_{ion}$ is the local ionic current density and can be written as $i_{ion} = \sigma_{ion}\nabla\phi$ (Ohm's law) inside the electrodes, where $\sigma_{ion}$ is electrolyte conductivity in the electrode. We note that Ohm's law is valid for binary and symmetric electrolyte (where diffusive current vanishes). Similarly, ionic current density in the spacer is also given by $i_{ion} = \sigma_{ion}\nabla\phi$. See Trainham and Newman,[40] Biesheuvel and Bazant,[41] and Hemmatifar et al.[32] for similar formulations of porous electrode systems. In this work, we use conductivity of a propylene carbonate solution of tetraethylammonium tetrafluoroborate (TEA-TFB) at 25 °C reported by Tyunina et al.[42] To this end, we first interpolate conductivity of free solution from the concentration-conductivity data ($\sigma_{ion,\infty}$) and then correct it for tortuosity as $\sigma_{ion} = \sigma_{ion,\infty}/\tau$ using the Bruggeman relation.

The balance between ionic current density in the electrolyte ($i_{ion}$), electronic current density in electrode matrix ($i_{elec}$), and external current can be written as

$$p_e i_{ion} + (1-p_e) i_{elec} = i_{ext}, \quad (7)$$

where $i_{ext}$ is the applied (external) current. Conservation of current in the spacer region (where no charge storage takes place) requires that the ionic current density is spatially uniform, i.e. $\nabla \cdot i_{ion} = 0$. So, in the one-dimensional case, ionic current density in the spacer is simply

$$p_s i_{ion} = i_{ext}. \quad \text{(spacer)} \quad (8)$$

We adopt a Helmholtz EDL model and relate electrolyte and electrode matrix potentials as

$$\phi - \phi_e = \rho/C, \quad (9)$$

where $\phi_e$ is electrode matrix potential and $C$ is specific capacitance for the EDL (in units of Farads per electrode volume). Note that current density in the electrode follows Ohm's law, $i_{elec} = \sigma_{elec}\nabla\phi_e$, where $\sigma_{elec}$ is electronic conductivity of electrode matrix. This conductivity is representative of, for example, compacted activated carbon powder.[43] With these assumptions, we take divergence of eq 9 and combine it with eq 7 to express the current balance equation in terms of model variables $c$, $i_{ion}$, and $\rho$ as

$$\left(\frac{1}{\sigma_{ion}} + \frac{p_e}{(1-p_e)\sigma_{elec}}\right) i_{ion}$$
$$= \frac{i_{ext}}{(1-p_e)\sigma_{elec}} + \frac{1}{C}\nabla\rho. \quad \text{(electrode)} \quad (10)$$

We stress that eqs 8 and 10 respectively describe ionic current density in the spacer and electrodes. In the case of known applied current (either constant or time-varying $i_{ext}$), the set of eqs 4-6, 8, and 10 fully describe charge/discharge dynamics of the cell. In the case where external voltage (and not external current) is known,

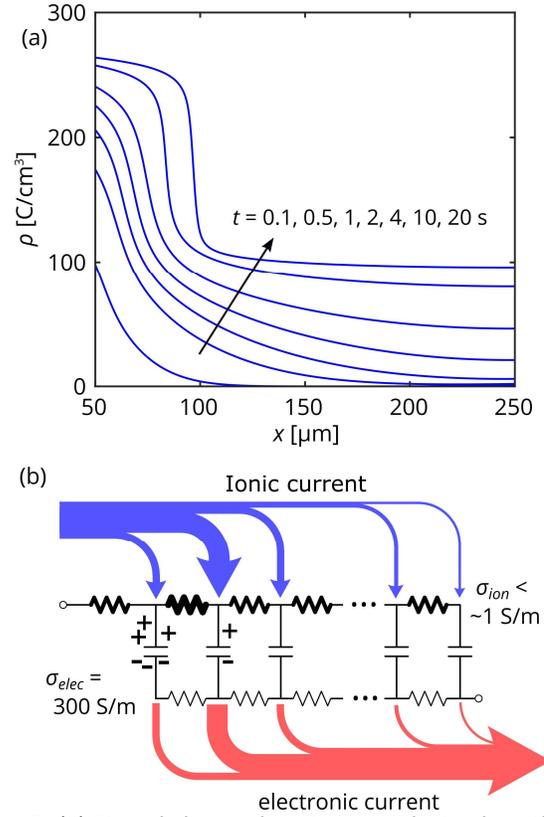

**Figure 2.** (a) Stored charge density in an electrode with high electrical conductivity (i.e. $\sigma_{elec} = 300$ S/m $\gg \sigma_{ion}$) as in traditional porous electrode systems versus position in a single electrode at various times during constant voltage charging at 2.7 V. (b) Schematic of the ionic and solid electrode transmission line models for porous electrode system. Ionic current entering from the spacer to the left quickly converts to electronic current to take advantage of the high electrode matrix conductivity ($\sigma_{elec} = 300$ S/m) compared to the solution in pores ($\sigma_{ion} \approx 1$ S/m). Consequently, charge is preferentially diverted to capacitive sites near the spacer. Near-spacer capacitive sites must be saturated before additional charge penetrates into electrode.

however, we need to enforce an extra condition to ensure consistency between external current and the resulting voltage. To this end, we set solution potential to zero at the midplane of the cell (symmetry plane), integrate the electric field along the cell (from $x = 0$ to $x = L_s/2 + L_e$), and use the potential equation ($\phi - \phi_e = \rho/C$) to arrive at

$$\frac{i_{ext}}{p_s}\int_0^{\frac{L_s}{2}} \frac{1}{\sigma_{ion}} dx + \int_{L_s}^{\frac{L_s}{2}+L_e} \frac{i_{ion}}{\sigma_{ion}} dx \quad (11)$$
$$-V_{ext}/2 = \tilde{\rho}/C,$$

where $V_{ext}$ is (constant or time-varying) external voltage, and $\tilde{\rho}$ is stored charge density evaluated at the electrode-current collector interface. Note that eq 11, at any given time, is linear in $i_{ext}$ and $V_{ext}$, and so the formulation is reminiscent of Ohm's law for an ideal resistor. Boundary and interface conditions are as follows:
(1) zero mass flux and ionic current at electrode-current collector interface ($\nabla c = 0$ and $\boldsymbol{i}_{ion} = 0$)
(2) symmetry in concentration ($\nabla c = 0$ at the midplane)
(3) continuity of concentration
(4) continuity of mass flux ($p_s D_s \nabla c|_s = p_e D_e \nabla c|_e$) and ionic current ($p_s \boldsymbol{i}_{ion}|_s = p_e \boldsymbol{i}_{ion}|_e$) at spacer-electrode interface

We here focus mostly on constant voltage charging, but note that high rate constant current charging shows similar effects. We implement this model in COMSOL Multiphysics (COMSOL Inc., Burlington, MA) using the equation-based modeling interface. We simulate a $A_e = 10 \text{ cm}^2$ area cell with spacer thickness of $L_s = 100 \text{ μm}$ and electrode thickness of $L_e = 200 \text{ μm}$ (i.e., total cell thickness of 500 μm). Due to the symmetry of the model, we only solve for half of the cell. In all simulations, we consider electrode material with volumetric capacitance of 200 F/cm³, porosity of 0.8, tortuosity of unity, and 0.8 mol/L concentration of TEA/TFB in propylene carbonate.

## 5. Results and discussion
### 5.1. Electrolyte depletion in an electrode with uniformly high conductivity (traditional design)

The local spatial distribution of electrolyte depletion depends strongly on the electrode conductivity. Common electrode materials traditionally have conductivities much higher than those of electrolyte solutions. The high conductivity of the electrode favors flow of electronic current in the electrode matrix rather than ionic current in the pore space. See Figure S1 of the Supplementary Information (SI) for the partition of current between the pore space versus the electrode matrix. Figure 2a shows stored charge density as a function of depth in the electrode simulated for constant voltage charging at 2.7 V with an electrode matrix conductivity $\sigma_{elec} = 300 \text{ S/m}$. The high (electrical) conductivity of the electrode compared to the solution causes preferential charging of the electrode near the spacer interface. The lowest resistance path for current is conversion from ionic to electronic flux as early as possible, with charging of the "front" of the electrode near the spacer as shown schematically in Figure 2b. Only when the electrode is highly charged locally and a significant local potential difference builds up between the electrode surface and solution, does the ionic current propagate deeper into the electrode.

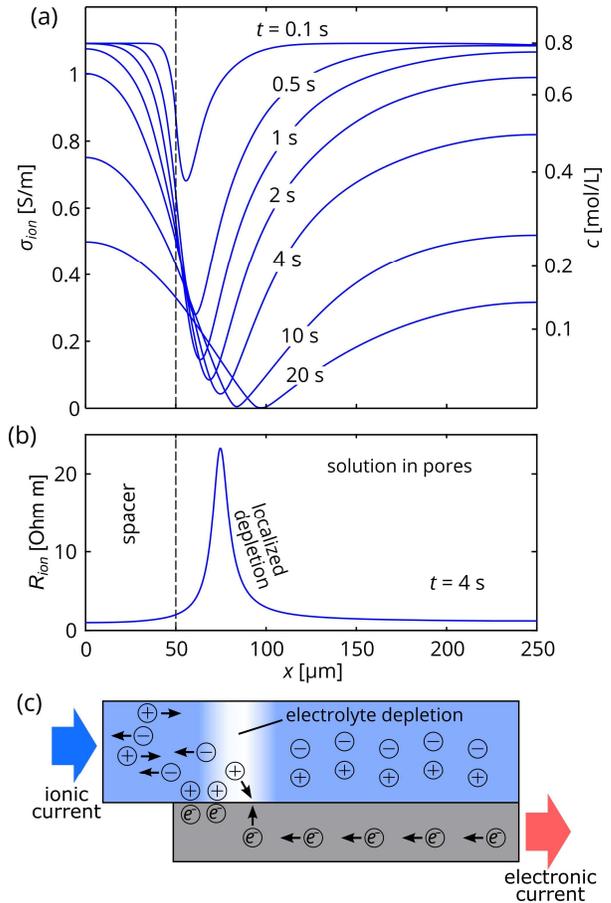

**Figure 3**. Electrolyte depletion in an electrode with high matrix conductivity (i.e. $\sigma_{elec} \gg \sigma_{ion}$) (a) Conductivity (left ordinate) and concentration (right ordinate) of electrolyte versus position in half of cell (symmetry line at $x = 0$) at various times during constant voltage charging at 2.7 V for electrode with high conductivity ($\sigma_{elec} = 300 \text{ S/m}$). Vertical dashed line is the spacer-to-electrode interface. (b) Solution resistivity profile at $t = 4 \text{ s}$ showing localized depletion near spacer/electrode interface. (c) Schematic of "front-to-back" charging. High electrode conductivity leads to localized charging near electrode/spacer interface. Resulting depletion creates a high resistance barrier that impedes ionic current.

We here show that the conventional design strategy of maximizing electrode conductivity can cause local depletion and a self-imposed starvation of the electrode. Note, the charging at the spacer/electrode interface corresponds to localized depletion of the electrolyte in this region. Figure 3a shows the effect of high electrode conductivity on depletion throughout the cell at various times. Other parameters used here are identical to those in Figure 2. The non-uniform charging creates a "valley" in concentration and conductivity near the entrance of the electrode. The local volume depleted of electrolyte is a high resistance in series with the rest of the electrode as shown in Figure 3b. The depletion region forms a high resistance barrier. The electrode regions to the right of this barrier are high in electrolyte concentration but the electrolyte is isolated and "trapped" within the deeper regions of the electrode. This impedes ionic current which must access the rest of the electrode (where electrolyte remains plentiful) to continue the charging process (Figure 3c).

### 5.2. Comparison case of low electrode conductivity (also resulting in non-uniform charging)

Before exploring optimal configurations of conductivity profiles, we consider the useful comparison case of an electrode with (electrical) conductivity uniformly lower than the initial electrolyte (ionic) conductivity. We show model results for such a case in Figure 4. Here, the path for ionic conduction through the electrode pore space offers lower resistance than electronic conduction through the electrode material. Ionic current is then driven through the electrode to the near-terminal region on the right (the "back" of the electrode as in Figures 4a and 4c) before it is converted to electronic current by adsorption into the double layer (Figure 4c). This extreme case results in a depletion region beginning at the rear of the electrode which grows with an interface moving toward the entrance (toward the left) as charging progresses. Figure 4a shows solution conductivity and concentration versus depth within one electrode at various charging times for electrode conductivity of $1 \text{ S/m}$ (compare to initial solution conductivity of $1.09 \text{ S/m}$). For this case, the ionic current never experiences a region of high resistance as it is always moving through less depleted regions with the depleted regions existing where charging has already completed (Figure 4c). The electronic current, however, always experiences the now significant (overly high and uniform value of) resistance of the electrode.

### 5.3. Proposed electrode with reduced and non-uniform conductivity for spatially uniform depletion

We here present our proposed design of a porous electrode with non-uniform and decreased values of conductivity to achieve approximately (spatially) uniform charging and electrolyte depletion. As we showed in Figures 2, 3, and 4, the electrode conductivity is essential in governing the spatiotemporal response of the ionic current and the resulting depletion. Here we consider modifications to the electrode conductivity profile to achieve more uniform depletion throughout the electrode. It is important to note that equal ionic and electrode conductivity does not lead to uniform depletion (this condition instead results in local depletion zones forming at the front and rear of the electrode, propagating rearward and frontward, respectively). Instead, a spatially varying electrode conductivity is required to produce spatially uniform depletion.

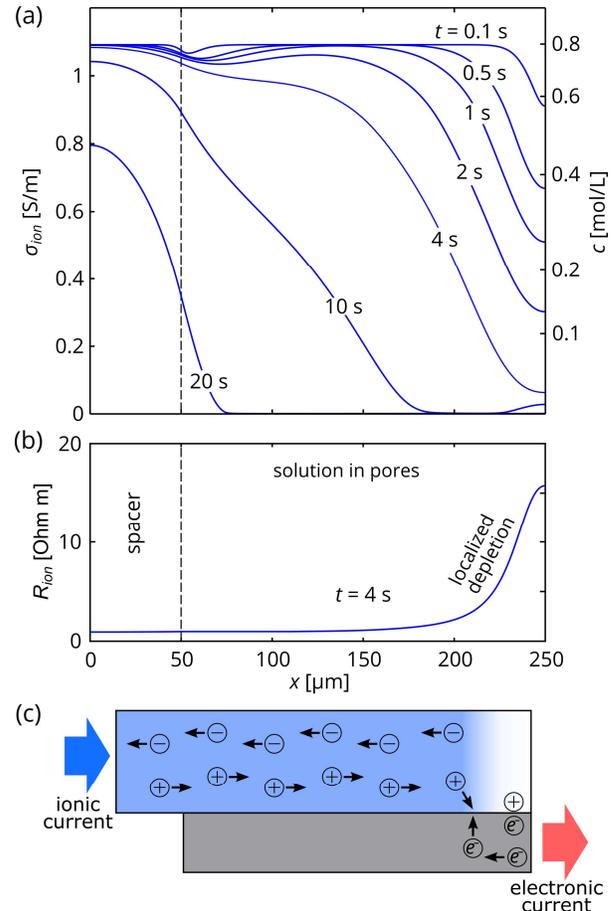

**Figure 4.** Electrolyte depletion in an electrode with uniformly very low conductivity (i.e. $\sigma_{elec} < \sigma_{ion}$). (a) Conductivity (left ordinate) and concentration (right ordinate) of electrolyte versus position in half of cell at various times during constant voltage charging at 2.7 V for electrode with low conductivity ($\sigma_{elec} = 1 \text{ S/m}$). (b) Solution resistivity profile at $t = 4 \text{ s}$ showing localized depletion near electrode/current collector interface. (c) Schematic of "back-to-front" charging. Low electrode conductivity leads to localized charging at the rear of the electrode near the electrode/current collector interface. Resulting depletion creates a high resistance region in the solution, but the ionic current never has to traverse this region. The electronic current in the matrix however experiences a uniformly high resistance.

We derive this analytically from our porous electrode model capturing non-uniform electrode conductivity profiles. From eq 6, the local rate of depletion is proportional to the divergence of the ionic current at all points in the electrode ($\partial \rho / \partial t = p_e \nabla \cdot i_{ion}$). A uniform divergence of ionic current in the electrode (i.e. a linearly varying ionic current) implies a uniform rate of depletion at that instant in time. For uniform depletion rate in the volume-averaged one-dimensional model considered here, we then require

$$\boldsymbol{i}_{ion} = \frac{1}{p_e}(1-\xi)\boldsymbol{i}_{ext}, \quad (12)$$

where $\xi = (x - L_s/2)/L_e$ is a dimensionless parameter representing location (depth) into the electrode. ($\xi = 0$ is at the electrode-solution interface, and $\xi = 1$ is at the electrode/current-collector interface). Substituting eq 12 into eq 10 and considering a uniformly charged electrode state ($\nabla\rho = 0$; i.e. $\rho$ constant everywhere), we derive a relationship between the required spatial dependence of electrode conductivity and solution conductivity to produce instantaneously uniform depletion

$$\sigma_{elec} = \frac{\xi}{1-\xi}\frac{p_e}{1-p_e}\sigma_{ion}, \quad (13)$$

which is a hyperbola in $\xi$. For this state, there is no dependence on external current and the required electrode conductivity distribution depends solely on the solution conductivity distribution via eq 13. To produce exactly uniform depletion over a finite time, the electrode conductivity must be time varying. However, as we show here, a time-invariant electrode conductivity distribution chosen to match the solution conductivity at a moderate level of depletion can provide a highly spatially uniform depletion rate over a large range of concentration change.

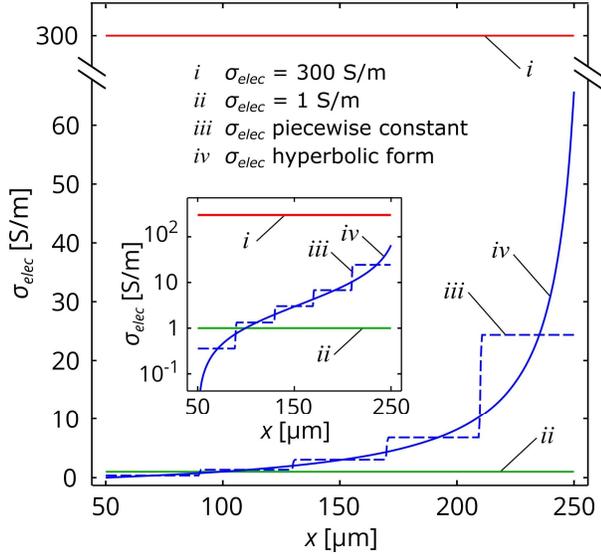

**Figure 5.** Electrode conductivity distributions considered in this work. The uniformly high conductivity (traditional) and uniformly low conductivity electrodes are denoted as $i$ and $ii$, respectively. The two versions of our proposed porous electrode with piecewise constant and hyperbolic form electrode conductivities are denoted as $iii$ and $iv$. The curve of case $iv$ is an analytically derived function for the electrode conductivity distribution leading to uniform electrolyte removal. The curve of case $iii$ is a simplified, discrete representation of this theoretical curve. The inset displays conductivity on a logarithmic scale to more clearly show small changes in conductivity near the spacer.

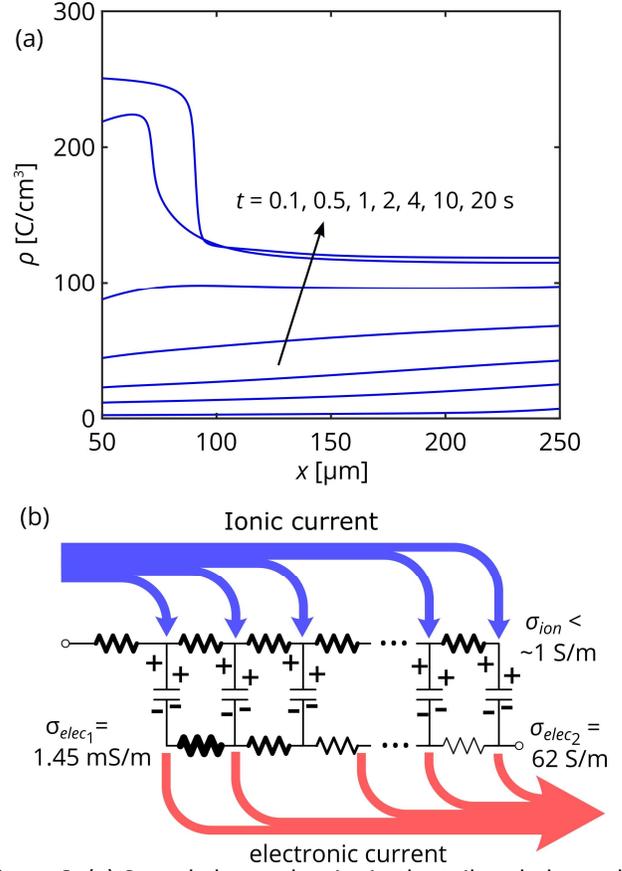

**Figure 6.** (a) Stored charge density in the tailored electrode with hyperbolic form conductivity (eq 13) versus position in a single electrode at various times during charging at 2.7 V. The charge density remains highly uniform until late times, where complete electrolyte depletion in the electrode leads to an increase in non-uniformity. (b) Schematic transmission line model for porous electrode with varying conductivity. Electrode matrix conductivity is limited to 1.45 mS/m and 62 S/m for the lower and upper bounds, respectively.

Figure 5 shows the electrode conductivity distributions considered in this work: a (traditional) high conductivity electrode (case $i$), a low conductivity electrode (case $ii$), an electrode with discrete conductivity values (case $iii$), and a continuously spatially varying electrode as given by eq 13 (case $iv$). The latter case is parameterized to produce a perfectly uniform depletion rate in our cell for a state corresponding to uniform charging and a uniform solution concentration of 0.445 mol/L (0.76 S/m), i.e. after ~30% depletion. For case $iv$, we limited the conductivity of the electrode to 1.45 mS/m and 62 S/m at the front and back of the electrode, respectively, to avoid numerical instabilities due to excessively low or high conductivity. Case $iii$ is a simple discrete representation of the theoretical curve of case $iv$ for ease in manufacturing. The electrode in this case consists of five segments with constant conductivity equal to the mean value of the continuously varying distribution at each segment midpoint (0.36, 1.33, 3.0, 6.8, and 24.4 S/m).

Figure 6 shows the spatiotemporal charging dynamics for our hyperbolic form conductivity profile (case $iv$) designed to charge the electrode more uniformly (and to avoid local depletion zones). In Figure S1 of the SI, we show this conductivity profile results in nearly linear current distributions over a large range of times which result in uniform charging of the electrode. The resulting charge distribution in Figure 6a is significantly more uniform than either the high electrode conductivity (Figure 3) or low electrode conductivity case (Figure 4) over these times. At later times (e.g. $> 10$ s), charging becomes less uniform with preferential charging at the electrode/spacer interface. This effect is the result of essentially complete electrolyte depletion in the electrode, as we will see in Figure 7. Additionally, we show a schematic of the transmission line model for the electrode with varying conductivity in Figure 6b. Note how a progressive increase of conductivity of electrode matrix with depth evenly distributes charging current.

Note, the electrode conductivity distribution was parameterized to generate uniform removal rate from a solution with an instantaneous uniform conductivity of 0.76 S/m (0.45 mol/L concentration). The initial solution for this case is somewhat more conductive (1.09 S/m). Consequently, current paths leading to charging at the rear of the electrode are initially slightly favored, but the charging profiles remain quite uniform compared to the single valued (uniform) electrode conductivity cases. Between $t = 2$ s and 5 s, the conductivity of the solution is sufficiently reduced that current paths leading to charging of the front of the electrode are then slightly favored resulting in a highly uniform state of charge at e.g. $t = 4$ s.

The relatively uniform charging corresponds to uniform reduction in electrolyte concentration throughout the electrode as charging proceeds (Figure 7a). This uniformity forestalls the formation of any high resistance regions in the solution permeating the pore space as seen in Figure 7b. Ionic current is allowed to flow through the entire depth of the electrode without significant impediment (Figure 7c). Continuously increasing electrode matrix conductivity gradually converts ionic current entering from the spacer to the right into electronic current in the matrix and charges the electrode uniformly. The resulting uniform and reduced depletion prevents the creation of high resistance regions in the solution while also not unnecessarily impeding the electronic current in the electrode matrix.

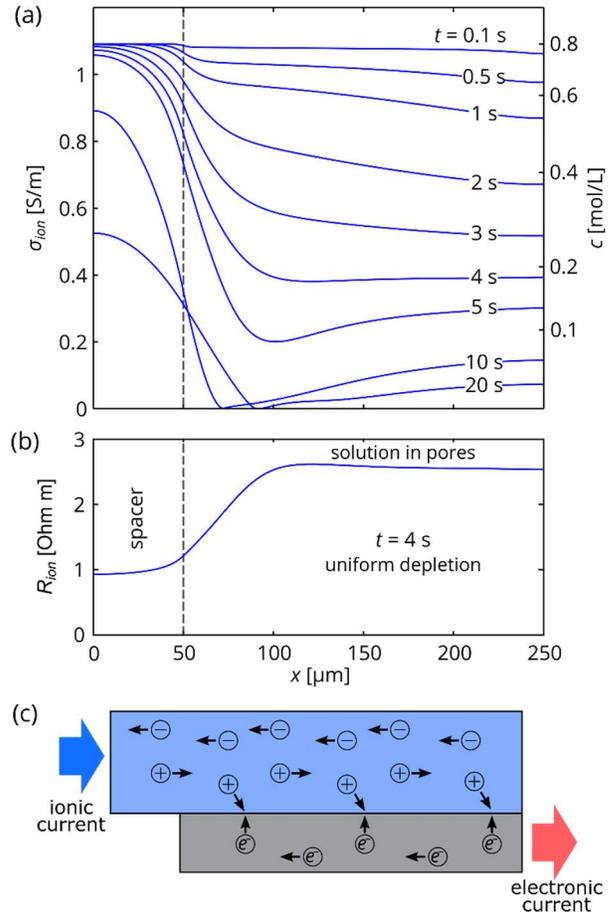

**Figure 7.** Electrolyte depletion for the continuously variable electrode conductivity shown in Figure 5. (a) Conductivity (left ordinate) and concentration (right ordinate) of electrolyte versus position in half of cell at various times during constant voltage charging at 2.7 V for electrode with low conductivity near the spacer ($\sigma_{elec} = 1.45$ mS/m) and high conductivity near the current collector ($\sigma_{elec} = 62$ mS/m). (b) Solution resistivity profile at $t = 4$ s showing uniform and reduced depletion throughout the electrode. (c) Schematic of uniform charging.

One consequence of the elimination of the localized high resistance region corresponding to depletion is reduction of the magnitude of electric field in solution during charging. Our simulations show a dramatic reduction in the maximum electric field in solution (up to 7 ×) during much of the charging period for the tailored resistance electrode compared to the traditional high conductivity case (see Section S.4 and Figure S4 of the SI). Additionally, our results show that a smooth electrode conductivity distribution is not required to produce this highly uniform depletion. In Figure S2 of the SI, we show the electrode with piecewise constant conductivity gives very similar results. We attribute this to the effect of electrolyte diffusion and capacitive response of the electrode material, where both act to reduce spatial non-uniformities during depletion. Importantly, the stepped conductivity profile (rather than a hyperbolic form) should significantly facilitate the electrode fabrication process.

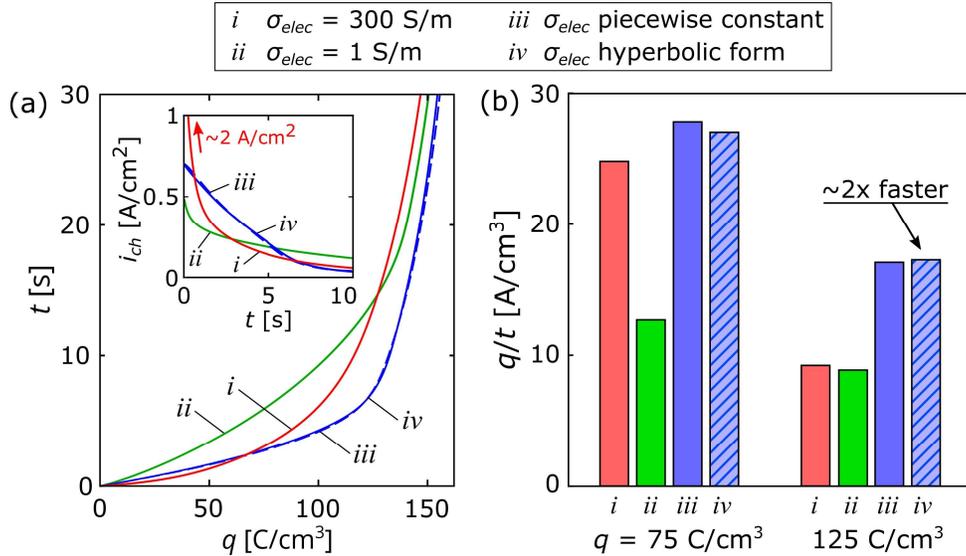

**Figure 8.** (a) Charging time to reach given charge state $q$ (charge normalized by single electrode volume of 0.2 cm$^3$) for the four electrode conductivity cases considered here. The inset shows corresponding charging current density ($i_{ch}$, current per cell surface area of 10 cm$^2$) versus time. Decrease of electrode conductivity decreases initial charging rates at very early times, but uniformly high conductivity electrodes (case $i$) can quickly develop depletion zones which subsequently strongly limit charging rate resulting in long charge times. (b) Average charging rate of each case at charge states of $q = 75$ and 125 C/cm$^3$. The variable conductivity electrode designs (cases $iii$ and $iv$) avoid depletion, and thus their charging rates quickly surpass the charging rate of the traditional electrodes (case $i$). At $q = 125$ C/cm$^3$ the average charging rate is 2-fold that of the traditional electrode.

In the next two sections, we will explore the effects of our proposed design for reduced and non-uniform electrode conductivity on charging time and energy consumption.

### 5.4. Effect of electrode conductivity magnitude and profile on charging rate of system

We here demonstrate increase in charging rate by tailoring electrode conductivity. The custom-tailored resistance is designed to avoid a local depletion, and this serves to improve overall (system volume averaged) charging rate. Note that decreasing the conductivity of the electrode is in contrast to traditional approaches which attempt to minimize all sources of resistance.[6] Figure 8a shows the time to reach a given state of charge for each of the electrode conductivity cases considered earlier (c.f. Figure 5). The inset gives the charging current density ($i_{ch}$, current per cell surface area of $A_e = 10$ cm$^2$) versus time. The traditional case (labeled $i$) initially shows the fastest charging but quickly ($< 1$ s) develops a strong depletion region near the spacer (c.f. Figure 3). This depletion significantly limits the rate of further charging, resulting in a rapidly decreasing charging current. The low and uniform conductivity case (labeled $ii$) shows charging which is always limited by the electrode resistance, with a slower but less variable charging rate. In contrast to these, the tailored electrode cases $iii$ and $iv$ start off with an intermediate charging current but quickly (at $t = 1$ s) overtakes the current of the traditional case $i$, as the more spatially uniform electrolyte concentration of cases $iii$ and $iv$ provides lower ionic resistance. As a result of these mechanisms, the tailored electrodes ($iii$ and $iv$) take a similar time as the traditional case $i$ to reach relatively low levels of charge (e.g. $q = 75$ C/cm$^3$) but take much shorter time to charge near capacity (e.g. $q = 125$ C/cm$^3$). Here, $q$ is the stored charge normalized by single electrode volume (0.2 cm$^3$). We also show the average charging rate, defined as cumulative charge per time, $q/t$, in Figure 8b for each case charged to $q = 75$ and 125 C/cm$^3$. At $q = 75$ C/cm$^3$, the average charging rate of tailored electrodes (cases $iii$ and $iv$) are shown to slightly exceed that of the two other cases.

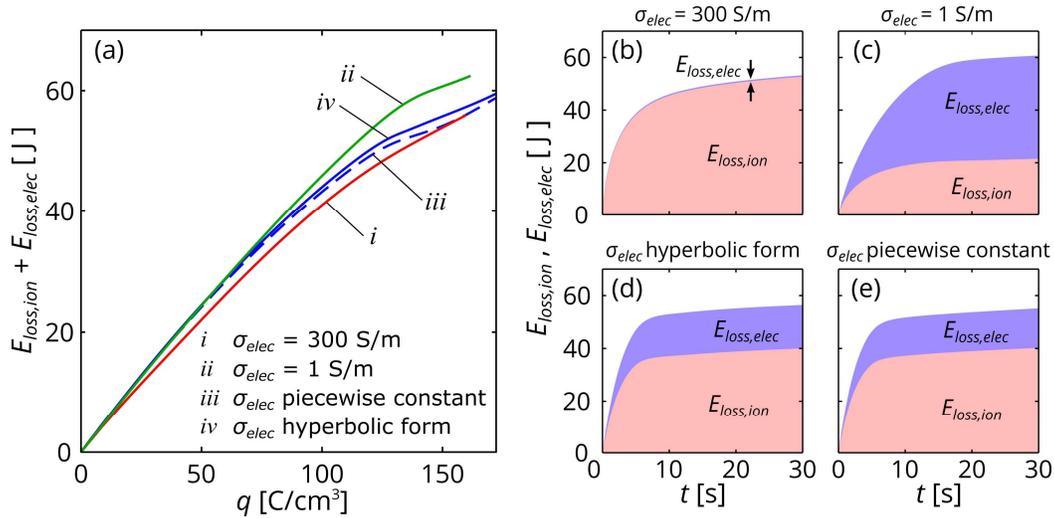

**Figure 9.** Cumulative resistive energy loss during charging at 2.7 V for the various electrode designs explored. (a) Total cumulative resistive loss for each of the four electrode conductivity cases versus stored charge. The (traditional) high conductivity electrode (case $i$) shows lowest total loss, but the non-uniform and lower conductivity electrodes (case $iii$ and $iv$) show $< 5\%$ additional loss. The spatially uniform, lower conductivity electrode (case $ii$) shows $< 15\%$ additional loss. (b)-(e) Electrode and solution energy loss ($E_{loss,elec}$ and $E_{loss,ion}$) contributions to resistive loss as a function of time. Increases in electrode resistive losses are largely offset by decreased solution resistive losses resulting from the decreased effect of localized depletion.

However, at $q = 125 \text{ C/cm}^3$, the average charging rate of the tailored electrodes is about 2 times that of the traditional electrode. This dramatic increase in charging rate is achieved with an overall lower conductivity of the electrode which is tailored to avoid the adverse effects of depletion near the spacer on charging dynamics of the system.

### 5.5. Effect of electrode conductivity magnitude and profile on energy loss

In traditional systems, most of the energy loss in the electrode region is due to low conductivity of the electrolyte (compared to the electrode). Hence, it is important to explore the effect of decreasing electrode conductivity to achieve more uniform and faster charging. Figure 9 shows plots of cumulative resistive energy loss during charging at 2.7 V for each of the four electrode conductivity cases of Figure 5. In Figure 9a, we show total cumulative resistive loss versus stored charge. The high conductivity electrode shows the lowest loss. However, the variable conductivity electrodes show only slightly higher losses. For example, for the cases we explored, variable conductivity electrode energy losses never exceed 5% additional loss compared to the high conductivity case and return close to the high conductivity loss value at the highest stored charges where depletion is most severe. The low conductivity electrode shows the highest loss, but still $< 15\%$ greater than the high conductivity case despite having an initial characteristic cell resistance which is $\sim 100\%$ larger. Figure 9b shows the contributions of the solution, $E_{loss,ion}$, and electrode, $E_{loss,elec}$, to resistive loss versus time, as given by

$$E_{loss,ion}(t) = 2A_e \int_0^t \left[ p_s \int_0^{L_s/2} \frac{i_{ion}^2}{\sigma_{ion}} dx + p_e \int_{L_s/2}^{L_s/2+L_e} \frac{i_{ion}^2}{\sigma_{ion}} dx \right] dt' , \qquad (14)$$

$$E_{loss,elec}(t) = 2A_e(1-p_e) \int_0^t \int_{L_s/2}^{L_s/2+L_e} \frac{i_{elec}^2}{\sigma_{elec}} dx \, dt' . \qquad (15)$$

The high conductivity electrode energy dissipation is dominated by solution loss since the resistance of the electrode is minimal. All electrodes with decreased conductivity show higher electrode losses as expected, but also show significantly reduced solution resistance losses compared to the high conductivity electrode due to the suppression of depletion at the electrode-spacer interface. The net effect is a minimal increase in resistive losses for electrodes with decreased conductivity (e.g. as discussed above, $< 5\%$ increase in energy loss for variable conductivity electrodes).

We here discuss energy loss in discharge phase, during which both electrolyte concentration and solution conductivity increase. Consequently, for discharge, the problem of electrolyte depletion does not exist and the total loss depends on the electrode conductivity and rate of discharge. In Figure S3 of the SI, we show resistive energy loss for complete discharge (to 0 V cell voltage) from $q = 150 \text{ C/cm}^3$ charge versus discharge rate. Our results show that, during discharge, the resistive loss for tailored electrodes is in general higher than that of conventional electrodes (although, at low discharge rates, the additional energy loss during discharge is negligibly small). Tailoring electrode conductivity to reduce localized depletion is, therefore, most applicable in situations with rapid charging and slower discharging. Such asymmetry is common in many electrochemical energy storage applications such as electric vehicles.[44] Additionally, we note that a number of battery chemistries (e.g. Li-ion and Pb-acid) experience electrolyte depletion on discharge as well. In these cases, high rate discharge may likewise benefit from the tailoring of electrode resistance. Furthermore, the electrode conductivity need not be symmetric on charge and discharge. For example, we hypothesize that the addition of a rectifying capability in the electrode could largely remove the additional loss on discharge and reduce discharge time constant while retaining the desired resistivity gradients during charge.

## 6. Conclusions

Depletion of electrolyte in electrochemical systems can have dramatic effects on their charging time responses and contribute to energy loss as well as system lifetime reduction. Maximization of electrode conductivity, as in traditional porous electrodes, can minimize energy loss, but also promotes highly localized depletion and electrolyte starvation of the electrode. We proposed a new approach wherein we reduce and control the distribution of matrix conductivity in porous electrodes as a means to avoid ion depletion and achieve highly uniform charging of the electrode. This can be used to improve charging response of electrochemical systems such as supercapacitors via the counterintuitive approach of increasing electrode resistance.

We presented a transmission line analogy useful in describing the principle of spatially non-uniform resistance electrodes to achieve uniform charging. Further, we developed a porous electrode theory transport model which captures the effect of electrode conductivity magnitude and distributions. We used this model to show that spatially tailoring of the electrode conductivity is required to produce uniform depletion. We developed an analytical expression for a time-invariant distribution of electrode conductivity that achieves largely uniform charging. We also presented a piecewise constant function which approximates the behavior of this idealized distribution and captures most of its benefit. We showed the reduction in localized electrolyte depletion achieved by the non-uniform electrode conductivity distribution can result in 2-fold increase in average charging rate. Using the porous electrode model, we showed that reductions in electrode conductivity do contribute to resistive loss, as expected, but this loss is largely counterbalanced by the decreased resistive loss in solution corresponding to depletion. A penalty in energy efficiency and response time is also paid during discharge for the reduced conductivity in the electrode. However, as discharge rate is decreased, this loss becomes negligibly small.

Here we modeled systems representative of supercapacitors for high rate energy storage, but this approach has broad application to many electrochemical systems. As an example, CDI systems are particularly susceptible to localized depletion effects.[30] Improved uniformity of depletion can likely enhance throughput of these systems. Furthermore, improvement of depletion uniformity using electrode resistance tailoring may provide a mechanism to combat the high field conditions associated with dendrite growth in certain battery chemistries.[26]

**Acknowledgments**
J.G.S., J.W.P, and A.H. gratefully acknowledge support from TomKat Center for Sustainable Energy at Stanford University. A.H. also gratefully acknowledges the support from the Stanford Graduate Fellowship program of Stanford University.

# Tailored Porous Electrode Resistance for Controlling Electrolyte Depletion and Improving Charging Response in Electrochemical Systems

James W. Palko,[a,b,‡] Ali Hemmatifar,[a,‡] and Juan G. Santiago[a,*]

‡ contributed equally to this work

[a] Department of Mechanical Engineering, Stanford University, Stanford, CA 94305, USA
[b] Department of Mechanical Engineering, University of California, Merced, CA 95343, USA

**Abstract**
This document contains supplementary information and figures further discussing ionic and electronic currents in our tailored electrode design, charging response of the piecewise constant conductivity design, energy loss considerations during discharge, and electric field distribution during charging.

## S.1 Effect of electrode conductivity on partition of electronic and ionic currents

Figure S1 shows the division of current between the pore space (ionic current) and the electrode matrix (electronic current) as a function of depth into the electrode for charging at 2.7 V. All other parameters are identical to those of Figures 2 and 6 of the main text. Figures S1a and S1b correspond to the conventional electrode (with $\sigma_{elec} = 300\,\text{S/m}$) and the hyperbolic-form conductivity electrode (tailored electrode), respectively. Unlike the conventional electrode, the tailored conductivity electrode results in nearly linear ionic and electronic current distributions over a large range of times. The divergence of these current distributions is thus quite uniform with position which results in uniform charging of the electrode.

## S.2 Charging response of piecewise constant electrode conductivity

Diffusion of electrolyte in solution and the capacitive response of the electrode material both act to suppress spatial non-uniformities during depletion. As a result, a smooth electrode conductivity distribution is not required to produce highly uniform depletion. In Figure S2, we show electrolyte depletion for an electrode with the piecewise constant, "stairstep" conductivity distribution shown in Figure 5 of the main text. At early times ($< 1\,\text{s}$), there are small spatial fluctuations in depletion corresponding to the abrupt variations in conductivity, but these rapidly decay to produce conductivity profiles indistinguishable from the electrode with smoothly varying conductivity. We expect the stepped conductivity profile to provide for significantly easier fabrication of the electrode.

## S.3 Energy loss of discharge process

During discharge, electrolyte concentration and solution conductivity increase. We expect control of the ion release process and distribution to have a much smaller effect on discharge energy loss than electrolyte depletion during charging. For electrodes with lower conductivity, we also expect larger discharge losses associated with the low electrode conductivity. However, in all cases, the total loss depends on the rate of discharge.

Figure S3 shows resistive energy loss for complete discharge (to 0 V cell voltage) from $q = 150\,\text{C/cm}^3$ charge versus discharge rate. As expected, the electrodes with decreased conductivity show greater loss compared to the uniformly high conductivity case. However, at lower discharge rate (lower discharge current density $i_{disc}$), the additional energy loss during discharge is negligibly small. Tailoring electrode conductivity to reduce localized depletion is, therefore, most applicable in situations with rapid charging and slower discharging.

As mentioned in the main text, such asymmetry is common in many electrochemical energy storage applications such as electric vehicles.[1] For comparison, with the cell considered, discharge times of about 2.3, 10, and 50 min (200, 50, and 10 mA discharge currents) correspond to 4.6 ×, 20 × and 100 × slower discharge rates, respectively, compared to the constant voltage charging time of about 30 s (see Figure 8 of the main text). At these rates, energy losses on discharge for hyperbolic form conductivity electrode (case $iv$ in Figure 9 of the main text), are only 8%, 2%, and 0.4% of energy loss at charging, respectively. We also note that a number of battery chemistries (e.g. Li-ion, Pb-acid) experience electrolyte depletion on discharge as well. In these cases, high rate discharge may likewise benefit from the tailoring of electrode resistance.

* To whom correspondence should be addressed. E-mail: juan.santiago@stanford.edu

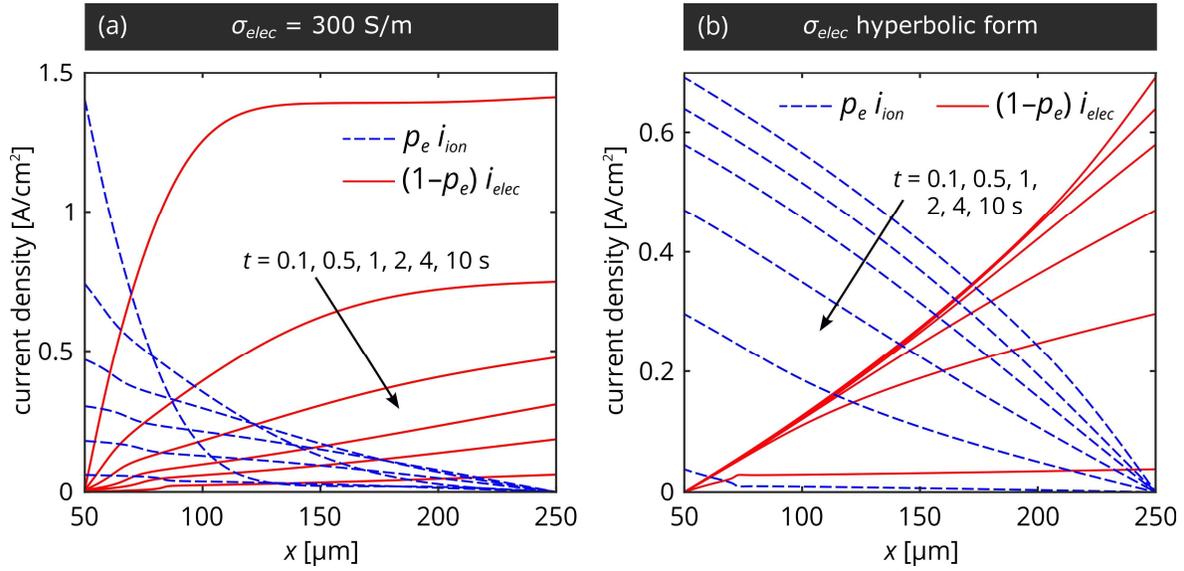

**Figure S1.** Current flow in (a) an electrode with high electrical conductivity ($\sigma_{elec} = 300$ S/m $\gg \sigma_{ion}$) as in traditional porous electrode systems and (b) a tailored electrode with hyperbolic distribution of conductivity. Shown is the ionic current in pore space (dashed blue curves) and electronic current in electrode matrix (solid red curves) versus position at various times during constant voltage charging at 2.7 V. In (b), nearly linear current profiles (and hence nearly uniform divergence of current) indicate uniform charging of the electrode. We show plots of stored charge distribution versus time in Figures 3 and 7 of the main text.

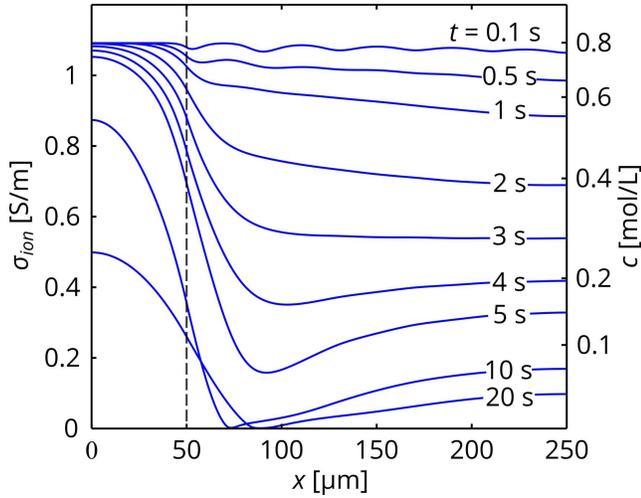

**Figure S2.** Conductivity of electrolyte versus position in half the cell at various times for piecewise continuous electrode conductivity distribution of Figure 5 of the main text. The simple piecewise distribution closely approximates the effect of the continuously varying conductivity profile but provides easier routes for manufacture.

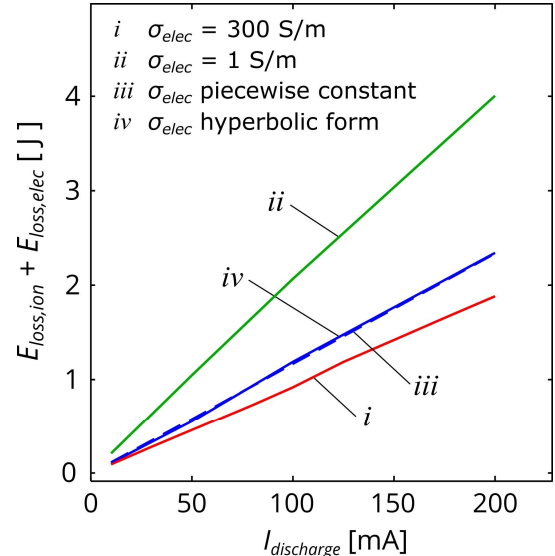

**Figure S3.** Resistive energy loss during complete constant current discharge (to 0 V cell voltage) versus discharge rate for all electrode conductivity cases. Reduced electrode conductivity corresponds to greater dissipation, but loss during discharge is strongly mitigated by slower discharge rates.

### S.4 Electric field distribution during charging

The localized charging and electrolyte depletion created by the traditional high conductivity electrode also leads to large, highly concentrated electric fields. Figure S4 shows the electric fields in solution permeating the electrode at various times during charging at 2.7 V for the high conductivity and hyperbolic tailored conductivity cases. The confined region of high solution resistance due to electrolyte depletion in the traditional case produces a strong peak in electric field which corresponds to the depleted region throughout the charging period.

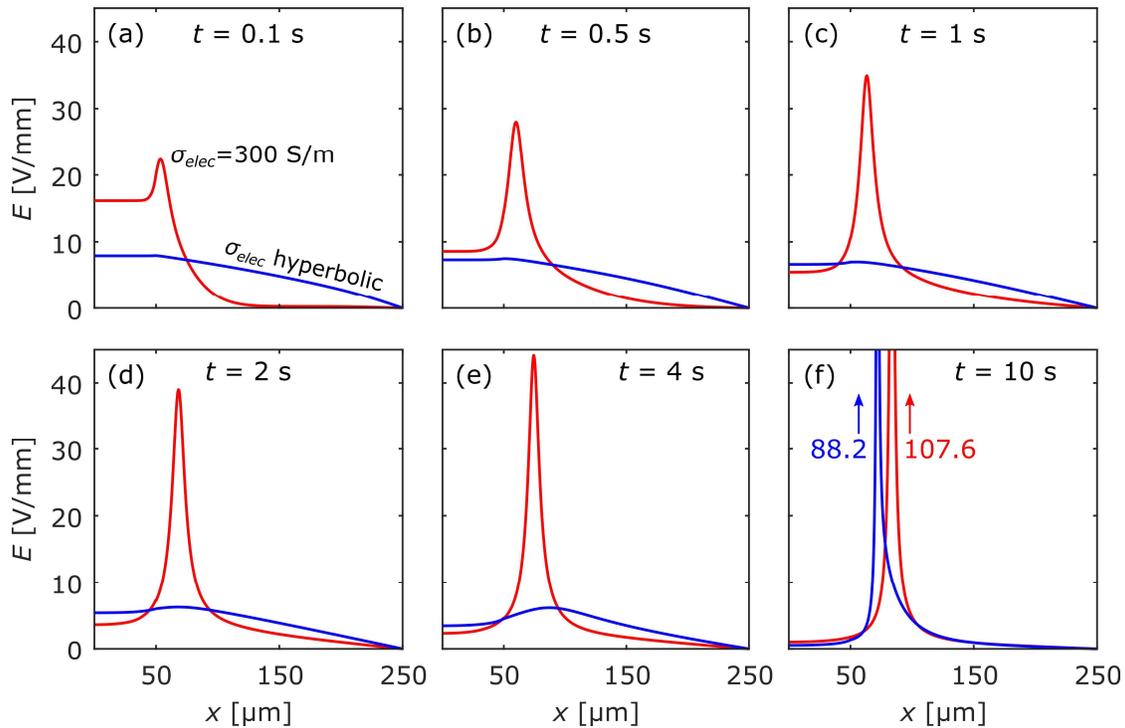

**Figure S4.** Calculated electric field in solution for high conductivity and hyperbolic tailored conductivity cases during charging of the cell at 2.7 V for times indicated. Localized electrolyte depletion created by the traditional high conductivity electrode leads to large, highly concentrated electric fields. The tailored electrode more uniformly removes electrolyte from solution producing less intense, more evenly distributed electric fields.

In contrast, the tailored electrode, which more uniformly removes the electrolyte from solution, does not produce a localized region of high resistance impeding the ionic current. Therefore, the voltage drop is more uniformly distributed throughout the electrode in this latter case, and the resulting electric field shows a much lower maximum. At late times (~10 s), once the electrolyte throughout the electrode has been mostly removed, the tailored electrode also shows the effect of depletion resulting in a peak in electric field but still with lower maximum value than the traditional case.

**References**

1 A. Vlad, N. Singh, J. Rolland, S. Melinte, P. M. Ajayan and J.-F. Gohy, *Sci. Rep.*, 2014, **4**, 928–935.